\documentclass[twocolumn,showpacs,preprintnumbers,amsmath,amssymb]{revtex4}
\usepackage{graphicx}
\usepackage{dcolumn}
\usepackage{bm}
\begin{document}
\newcommand{\W}{8cm}
\newcommand{\ud}{\rm d}
\newcommand{\un}{~\mathrm}
\newcommand{\ie}{{\em i.e. }}
\newcommand{\eg}{{\em e.g. }}
\newcommand{\unm}{~\mu\mathrm{m}}
\newcommand{\indice}[1]{\textnormal{\scriptsize{#1}}}

\preprint{APS/123-QED}
\title{Failure mechanisms and surface roughness statistics of fractured Fontainebleau sandstone}

\author{L. Ponson$^{1}$}\altaffiliation{Present address: COPPE/UFRJ Civil Engineering Program, Rio de Janeiro, RJ 21945-970, Brazil}\email{laurent.r.ponson@wanadoo.fr}
\author{H. Auradou$^1$}\email{auradou@fast.u-psud.fr}
\author{M. Pessel$^2$}
\author{V. Lazarus$^{1,3}$}
\author{J.P. Hulin$^1$}
\affiliation{$^1$ Laboratoire Fluide, Automatique et Syst{\`e}mes Thermiques,
UMR No. 7608, CNRS, Universit{\'e} Paris 6 and 11,
B{\^a}timent 502, Universit{\'e} Paris Sud, 91405 Orsay Cedex, France.\\
$^2$ Laboratoire Int{\'e}ractions et Dynamique des Environnements de
Surface, UMR 8148 I.D.E.S., CNRS-Universit{\'e} Paris-Sud 11,
B{\^a}timent 504, 91405 Orsay Cedex, France. \\
$^3$ Universit\'e Pierre et Marie Curie-Paris 6, CNRS,
UMR 7190 (Institut Jean Le Rond  d'Alembert), Boite 162, 4 place Jussieu, 75005 Paris, France.}
\begin{abstract}
In an effort to investigate the link between failure mechanisms and the geometry of
fractures of compacted grains materials, a detailed statistical analysis of the  surfaces of fractured
Fontainebleau sandstones has been achieved. The roughness of samples of different widths  $W$ is shown to
 be self affine with an exponent $\zeta=0.46\pm 0.05$
over a range of length scales ranging from the grain size $d$ up to
an upper cut-off length $\xi \simeq 0.15 W$. This low $\zeta$ value
 is in agreement with measurements  on other sandstones and on sintered materials.
The probability distributions $\pi_{\delta z}(\delta h)$ of the
variations of height over different distances $\delta z > d$ can be
collapsed onto a single Gaussian distribution with a suitable
normalisation and do not display multifractal features. The
roughness amplitude, as characterized by the height$-$height
correlation over fixed distances $\delta z$, does not depend on the
sample width, implying that no anomalous scaling of the type
reported for other materials is present. It is suggested, in
agreement with recent theoretical work, to explain these results by
the occurence of brittle fracture (instead of damage failure
in materials displaying a higher value of $\zeta \simeq 0.8$).
\end{abstract}

\maketitle
\section{Introduction}
The characterization of  the geometry of fracture surfaces  is of
considerable interest for a broad range of practical applications
ranging from the hydrology of fractured reservoirs~\cite{Adler,
Zimmerman} to the derivation of friction laws~\cite{Chakrabarti}.
Their statistical analysis is also relevant to the understanding of
the physics of fracture: The roughness may indeed be expected to
reveal failure mechanisms occurring at the microstructure
scale~\cite{Bouchaud4}. It is now well established that fracture
surfaces satisfy a scaling invariance known as {\it
self-affinity}~\cite{Bouchaud9, Maloy, Poon, Schmittbuhl8, Schmittbuhl5,
Ponson5}. A {\it{self-affine}} profile is one that is statistically
invariant under the scaling relation: $z\rightarrow \lambda z$ and
$h(z) \rightarrow \lambda^\zeta h(z)$ where the $z$ axis is
 in the mean fracture plane and normal to the direction of
crack propagation, $h(z)$ is the surface height profile and $\zeta$
is the self affine exponent. The scaling invariance implies that
the probability  $\pi_{\delta z}(\delta h)$ to observe a  variation  of height $\delta h$ over an horizontal distance $\delta z$ follow  the scaling law:
\begin{equation}
\lambda^{\zeta}\pi_{\lambda \delta z}(\lambda^\zeta \delta
h)=\pi_{\delta z}(\delta h) \label{eq:eq1}
\end{equation}
The prefactor $\lambda^\zeta$ results from the normalization of
$\pi_{\delta z}(\delta h)$. As a direct consequence, the
height$-$height correlation function $\Delta h(\delta z)$ satisfies:
\begin{equation}
\Delta h(\delta z) = \langle (h(z+\delta z) - h(z))^2
\rangle^{1/2}_z \propto \delta z^{\zeta} \label{eq:eq2}
\end{equation}
In many materials including rocks like granite and
basalt~\cite{Poon,Schmittbuhl8, Schmittbuhl5}, the roughness
exponent $\zeta$ was found to be close to $0.8$, suggesting at first
that it was a {\it universal} value~\cite{Bouchaud9, Maloy}.
However, different exponents  were reported for Berea sandstone
$\zeta \simeq 0.4-0.5$~\cite{Boffa} and then  on synthetic sandstone
made of sintered glass beads~\cite{Ponson6}. In this latter case,
the exponent $\zeta$ is equal to $0.40 \pm 0.04$ independent of the
porosity in the range investigated ($3\un{\%} <\phi<26\un{\%}$).
This difference raises important fundamental questions regarding the
physics of fracture, namely whether $\zeta$ is determined by the
structure of the material or by the
failure mechanism.

 In support of the second possibility, a recent work~\cite{Bonamy2}
suggests that  the value of $\zeta$ depends on  
the existence of a {\it process zone} where damage processes take place.
On the one hand,  at length scales smaller than the size of the
process zone, 
fracture surfaces would develop from the growth and coalescence 
of microdefaults (cracks or voids):  This  {\it damage} failure (also called
{\it quasi-brittle} failure) would lead to an exponent  $\zeta \simeq 0.8$.
 On the other hand, at length scales larger than the process zone, the fracture
surfaces would result from  the continuous propagation of a crack. For this {\it brittle} failure,
 the exponent would be $\zeta \simeq 0.4$, reflecting, at the continuum scale,
the toughness distribution of the microstructure scale.
Two-dimensional numerical simulations realized under these two types of assumptions
confirm these results by predicting lower $\zeta$ values for brittle fractures~\cite{Katzav} than
when damage is introduced~\cite{Bouchbinder}.

While most studies dedicated to the roughness of cracks dealt with surfaces
with $\zeta\simeq 0.8$~\cite{Bouchaud9, Maloy, Poon, Schmittbuhl8, Schmittbuhl5,
Ponson5,Lopez4,Morel,Mourot2,Santucci}, the present work investigates
these problems through an extensive
analysis of the statistical properties of the roughness of fractured
Fontainebleau sandstone for
which $\zeta \simeq 0.4-0.5$.

 In addition to the self-affine exponent $\zeta$, another important
statistical characteristic is the
$\pi$ distribution of the surface height variations: On granite
surfaces ($\zeta = 0.8$), it was found to be
Gaussian~\cite{Santucci} and a single self-affine
exponent was needed to describe the correlation of the surface
heights along the crack front direction.

Another important point is the dependence of the roughness
statistics on the size of the samples and on the distance to the
initiation zone. In the transient propagation region of fractures
initiated  from a straight notch, the amplitude of the roughness is
observed to depend at all length scales on the distance to the
initiation~\cite{Lopez4,Morel,Mourot2}. This variation may be
described by a so called {\it anomalous} scaling law involving
 a new exponent $\zeta_\indice{global}>\zeta$ referred as the {\it global}
roughness exponent. A consequence of anomalous scaling is expected
in regions far from the initiation and where the roughness is fully
developed~\cite{Morel,Mourot3}: The roughness amplitude in a window
of size $\delta z$ normal to the crack propagation  would then scale with the
width $W$ of the fracture surface as:
\begin{equation}
\Delta h(\delta z) \propto \delta z^\zeta
W^{\zeta_{global}-\zeta} \label{eq:eq3}
\end{equation}
 In the present study, a statistical analysis of the same type is
performed on  fractured Fontainebleau
sandstone samples of various widths $W$ (perpendicular to the direction of crack
propagation).  For each surface, the roughness exponent
$\zeta$ and the range of length scales over which a self-affine
description is valid are inferred from the variations of the
correlation function $\Delta h$ with $\delta z$. Their dependence on
the width $W$ of the samples and on the location on the fracture surface
is investigated in order to look for the anomalous scaling features observed on other materials. The probability distribution of the variations of height over given distances
$\delta z$ are also investigated, in order to evaluate the  magnitude of
multiscaling effects and their dependence on  $\delta z$.

The differences between these results and the
characteristics reported previously for other materials will be discussed and we shall seek to relate them to the failure  mechanisms taking place at scales larger
than the grain size in Fontainebleau sandstone.
\section{Sample characterization and experimental setup}
The X-Ray analysis of the Fontainebleau sandstone samples used in
the present work indicated that they contain  $99\un{\%}$ quartz.
The size of the grains range from $100\unm$ to $500\unm$ with a mean
value $d \simeq 270\unm$. $16$ SEM micrographs covering a total area
of $20\un{mm} \times 20\un{mm}$
 were used to measure the porosity.
The pictures are thresholded at a gray level adjusted so as to discriminate between grains and voids.
 The porosity, defined as the ratio between
 the void and total areas,
is found to be $10\pm 1\un{\%}$.\\
\begin{figure}
\includegraphics[width=\W]{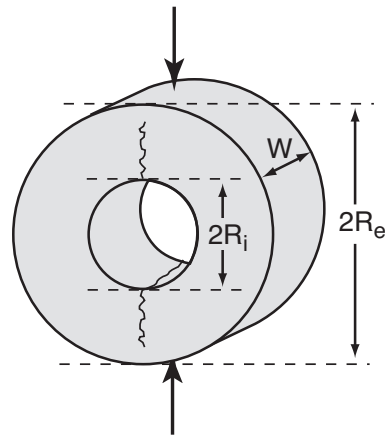}
\centering \caption{Schematic view of the toroidal sample (outside radius $60\un{mm}$,
inside radius $25\un{mm}$) and of forces  applied during the modified Brazilian test.}
 \label{fig:fig1}
\end{figure}
Slices of widths ranging from $W=7.8$ to $51.5\un{mm}$ were sawed off from
Fontainebleau paving stones.
Toroidal samples were obtained by using hole saws  to core out the
slices (see Fig.~\ref{fig:fig1}). This was achieved by first
piercing a circular hole with a $25\un{mm}$ diameter saw and then a
 second hole of diameter $60\un{mm}$ using  the same spinning
 axis. Four
slices of widths $W=7.8$, $14.1$, $26$ and $51.5\un{mm}$ were machined
and placed in a hydraulic press to perform modified Brazilian
fracture tests~\cite{Carneiro2}. The latter is achieved by applying
a uniaxial stress at two opposite points on the outside of the torus. The
load is increased until two symmetrical cracks propagate from the
central hole toward the outside where the compressive forces are applied.

\begin{figure}
\includegraphics[width=\W]{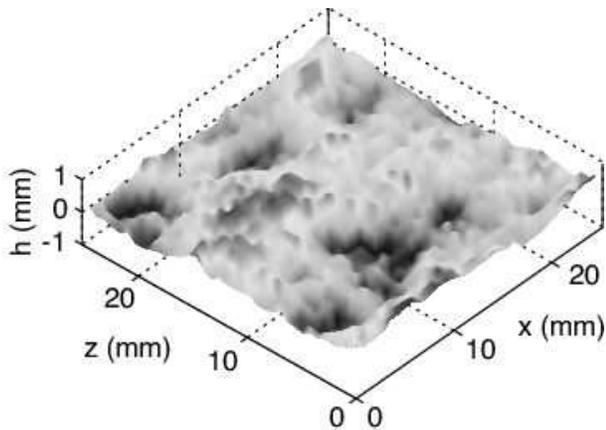}
\centering \caption{Topographic image ($248\times 252$ pixels) of a fractured sandstone
surface with a crack propagation along the $x$-axis.
Sample width:  $W=26\un{mm}$ (along the $z$-axis).}
 \label{fig:fig2}
\end{figure}

After the failure, one of the fractured blocks is secured to a
computer controlled horizontal displacement table. The samples are
moved stepwise using steps of lengths $\Delta z$ (See Tab.~\ref{tab:tab1}
for a detailed list of the actual values.) At each
new position, a sensing needle is displaced vertically by a computer
controlled microstep motor with a resolution of $\simeq 1\unm$. The
contact is detected by a vertical deviation of the needle. The
repeatability of the measurement, verified by scanning several times
the same profile, is $1\unm$. After the detection of the contact,
the sensor is raised by $200\unm$ before the sample is moved
laterally again. The sequence is repeated $n_z$ times. Then, the
sample is displaced in the perpendicular direction along the $x$-axis  by a
step of length $\Delta x$ and the scanning is repeated again.
Finally, one obtains surface maps including $n_z\times n_x$ points.
Figure~\ref{fig:fig2} shows one of these maps that displays troughs
and bumps with, for the largest, a typical amplitude of the order of
one millimeter. In the following, we discuss a detailed
statistical analysis of the spatial correlations of the roughness.\\
\begin{table}
\begin{ruledtabular}
\begin{tabular}{ccccccc}
$W$ & $\Delta z$ & $\Delta x$ & $n_z$ &$n_x$ &$\zeta$ &$\zeta_\indice{ft}$\\
\hline
$51.5$ & $50$ & $1000$ & $1007$ & $30$ & $0.51$ & $0.50$ \\
$26$ & $100$ & $100$ & $248$ & $252$ & $0.48$ & $0.46$ \\
$14.1$ & $50$ & $250$ & $263$ & $136$ & $0.46$ & $0.43$ \\
$7.8$ & $25$ & $250$ & $273$ & $133$ & $0.43$ & $0.45$ \\
\end{tabular}
\end{ruledtabular}
\caption{\label{tab:tab1}Characteristic parameters of fracture
surface geometry. $W$ ($\mathrm{mm}$): specimen width. $\Delta z$,
$\Delta x$ ($\mu\mathrm{m}$): measurement steps along the $z$ and
$x$. $n_z$ and $n_x$: number of recorded points along the $z$ and
$x$ directions. $\zeta$ and $\zeta_\indice{ft}$: self-affine
exponents of profiles oriented in the $z$ direction measured
respectively using  the correlation function and the Fourier power
spectrum.}
\end{table}
\section{Experimental results}
\subsection{Characterization of the surface roughness}
\label{sec:secA}
We first analyze profiles parallel to the $z$-axis, \ie
normal to the direction of crack propagation, and located far enough
from the initiation so that the roughness properties are
statistically stationary. The scaling properties of these profiles
are characterized by their 1D
height$-$height correlation function $\Delta h(\delta z)$ defined by Eq.~(\ref{eq:eq2}).

\begin{figure}[b]
\includegraphics[width=\W]{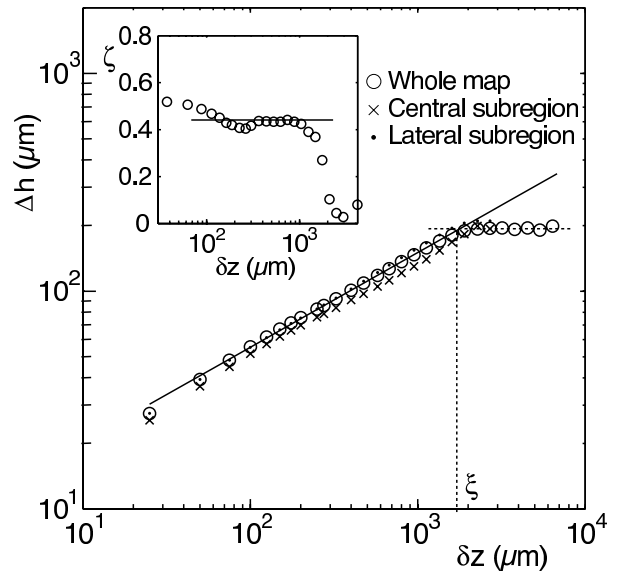}
\caption{Log-log representation of the height$-$height correlation
function $\Delta h$ as a function of $\delta z$ computed along the
crack front direction ($z$-axis) for a sample of width $W=7.8\un{mm}$.
Straight line: Linear regression of the data for $\delta z < \xi$ with slope $\zeta=0.43$.
Inset: Local slope of the
correlation function.} \label{fig:fig3}
\end{figure}

Figure~\ref{fig:fig3} represents in a log-log scale the variations of
$\Delta h$ as a function of the distance $\delta z$ for one of the
samples. Two distinct behaviors are observed: For $\delta z < \xi$,
$\Delta h$ varies as a power law of $\delta z$  (straight line in a
log-log  representation) while, for $\delta z > \xi$,  $\Delta h$
remains roughly constant ("plateau domain"). The crossover length
scale $\xi$ is, here, defined as the abscissa of the intersection
between the power law fit (solid line in Fig. \ref{fig:fig3}) and
the plateau variation (dotted horizontal line). The local slope of
the correlation function (see inset of Fig. \ref{fig:fig3}) is constant except for $\delta z$ values smaller than $\simeq 100\unm$: This
corresponds  roughly to the grain size $d$ which represents therefore a lower
boundary of the self affine domain. These results indicate  that, at
intermediate length scales, the profiles are
self-affine and characterized here by a roughness exponent $\zeta=0.43$.\\
The robustness of this self-affine description was verified by using
other statistical methods such as the Fourier analysis of the
profiles \cite{Feder}. The exponents obtained in this way are given
in Tab.~\ref{tab:tab1} and are globally independent of the method:
Regarding the dependence on $W$, a global increasing trend is
observed but the total variation is very weak,  less than  $0.1$, which is of the same 
order as the interval of confidence on the value of
one exponent estimated for a given width. The self-affine exponent
obtained after averaging over all data is $\zeta = 0.46\pm 0.05$.
Applying the same analysis to profiles parallel to the $x$
direction, \ie the direction of crack propagation, leads to a self-affine exponent
equal to  $0.49 \pm 0.05$. These observations are consistent with
measurements performed on other natural and
artificial sandstones \cite{Boffa, Ponson6, Bonamy2}.

For $\delta z$ larger than $\xi$, the variations of $\Delta h$ level off
indicating that the surface is flat at large scales. In order to test the
robustness of the value of $\xi$ with respect to possible bias introduced
 by the statistical analysis, the computation
was repeated on data sets including only one out of two data points.
The value of the cut-off length remained the same.
Finally,  the same procedure was applied in different
regions of the surface. Each of the surfaces was divided into three
regions of equal width $W/3$. Two of these subsurfaces are thus
on the sides of the sample while the third one covers the central
region. In Fig.~\ref{fig:fig3}, correlation
functions $\Delta h$ computed on the central ($ \times $) and on one of
the lateral ($ \cdot $) subregions are overlaid. All curves fall
on top of each other, showing that the crossover length $\xi$ is
independent of the region of the fracture surface. Therefore, $\xi$ is a relevant length
scale of the problem  which reflects neither an influence of  the sides of the sample
nor of  the sampling period of the surface maps.

Before studying the scaling behavior of $\xi$ with the sample width,
we analyze now the distribution $\pi_{\delta z}$ of the variations of height.

\subsection{Statistical distribution of the fluctuations of height}
\begin{figure}
\includegraphics[width=\W]{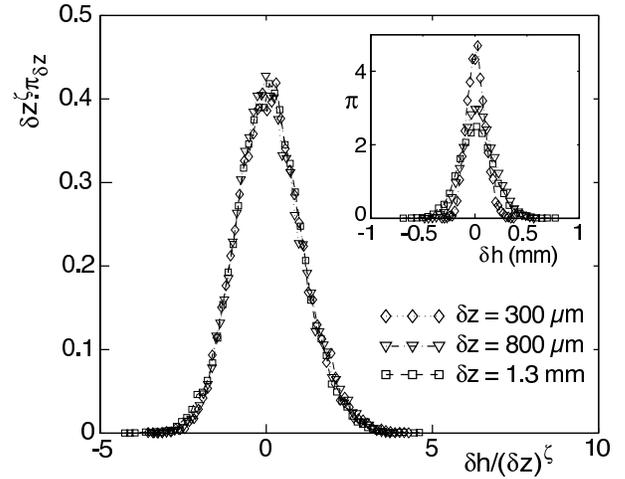}
\caption{Probability distribution of height variations over a
distance $\delta z$ normalized using Eq.(\ref{eq:eq5}) with $\zeta = 0.43$ as 
obtained by the analysis of the second moment $\Delta h$ of
the height distribution (See Sec. \ref{sec:secA}). $\delta h$ and $\delta z$
have been normalized by a same length scale length scale $\ell = 35 \unm$ chosen
 so that the second moment of the normalized distributions is equal to $1$. Inset : Plot
in linear coordinates of $\pi_{\delta z}(\delta h)$ for $300\unm \le
\delta z \le 1.3\un{mm}$.} \label{fig:fig4}
\end{figure}

For each sample, a scan including at least $30,000$ measurement
points has been performed. Such a large data set allows one to compute
with a good accuracy the probability distributions $\pi_{\delta z}(\delta h)$  for values of
$\delta z$ ranging from the typical grain size up to several times the grain diameter.
Fig.~\ref{fig:fig4} displays such  probability distributions corresponding to
several  $\delta z$ values ranging from $300\unm$ to $1.3\un{mm}$, i.e. larger
than the mean grain diameter $d$.

For self-affine profiles, one expects $\pi_{\delta z}$ to satisfy
Eq.~(\ref{eq:eq1}), valid for any scaling parameter $\lambda$. Using
the particular value $\lambda=1/\delta z$, this equation becomes :
\begin{equation}
\pi_{\delta z}(\delta h) = \left(\frac{1}{\delta z}\right)^\zeta
\pi\left(\frac{\delta h}{\delta z^\zeta}\right) \label{eq:eq5}
\end{equation}
Therefore, using the normalized variable $u = \delta h/\delta
z^\zeta$, all the distributions $\delta z^\zeta \pi_{\delta z}$
should collapse onto a same master curve $\pi(u)$, at least for
$\delta z$ values pertaining to the self-affine regime. Such a
collapse is indeed observed in Figure~\ref{fig:fig4}: The same data
as in  the insert have been plotted in these normalized coordinates
using the roughness exponent $\zeta = 0.43$ measured from the scaling
of the correlation function (see Fig. \ref{fig:fig3}).\\
One of the key
consequence of this  precise  collapse is that all moments
$\langle (h(z+\delta z) - h (z))^p \rangle_z^{1/p}$ of order $p>1$ of the $\pi_{\delta z}$
distributions must scale as $\delta z^\zeta$. This is a clear evidence of the lack of
multiscaling of the surface investigated.
In other words, the single exponent
$\zeta$ is enough to describe the full statistics of the variations of height on the
fracture surface along the $z$-axis, at least in the range of length scales $d < \delta z < \xi$.

\begin{figure}[b]
\includegraphics[width=\W]{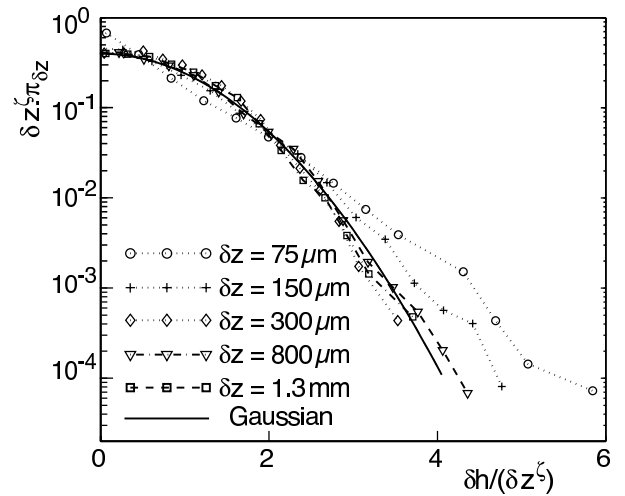}
\caption{Probability density $\pi_{\delta z}(u)$ in a
semi-logarithmic representation for $75\unm \le \delta z \le
1.3\un{mm}$. Here $u=\frac{\delta h}{\ell} \frac{\ell^\zeta}{\delta z^\zeta}$ with $\ell=35 \unm$
 and $\zeta = 0.43$ as obtained
  from the analysis of the second moment of the distribution (See
  Sec.\ref{sec:secA}). For clarity only positive $u$ are displayed.
  Continuous line : Gaussian distribution $\frac{1}{\sqrt{2\pi}}e^\frac{-x^2}{2}$.} \label{fig:fig5}
\end{figure}

Let us now focus on the master curve $\pi(u)$ which has the characteristic
 bell-like shape of Gaussian distributions. This is more precisely demonstrated
  in the semi-logarithmic plot of Fig.~\ref{fig:fig5} in which a Gaussian
  curve  (continous line) with a second moment equal to one is observed to
  coincide with the  experimental  data sets corresponding 
  to  $\delta z \ge 300\mu m$,
  {\it{i.e.}} $\delta z \ge d$.
  The statistical properties of the surface
 are therefore fully described  by this Gaussian distribution and by the single roughness exponent $\zeta = 0.43$.

We have also represented the $\pi_{\delta z}$-distributions
corresponding to length scales smaller than the mean grain diameter
$d = 270\unm$, namely  $\delta z = 75$ and $150\unm$. They clearly
do not collapse with the other curves although their second moment
was shown to scale roughly as $\delta z^\zeta$ in this range of
scales (see Fig. \ref{fig:fig3}). It means that, in the domain of
small distances ($\delta z < d$), the $\pi_{\delta z}$ distributions
are not Gaussian and their moments cannot be rescaled
 with a single exponent. In this range of $\delta z$ values,  it would be possible to rescale all the moments
  $\langle (h(z+\delta z) - h(z))^p \rangle^{1/p}$ by using 
  several exponents $\zeta_p$ (not shown here): The
relevance of such a multiscaling description at length scales for
 which the geometry of a single grain has
 a predominant effect is however questionable.

These results bring new insight to a recently debated question
concerning crack lines resulting from the rupture of paper sheets
that were found to exhibit multiscaling \cite{Bouchbinder2}, at
least at small scales, \ie at length scales comparable with the
length of the fibers of the paper. Other observations, performed on
the same material, but over a wider range of length scales, reported that,
at larger scales, there was
a crossover from this multiscaling behavior towards a self-affine
one~\cite{Santucci}. These results  may  have
similarities with  our experiments, even though they correspond to
the 3D rupture of a very different material: In our
 case, the crossover length correspond to the mean grain 
diameter $d$ of the sandstone specimen.

\subsection{Dependence of the statistical properties on the sample width}
In addition to the lower limit of the order of the grain diameter $d$ discussed above, the self-affine domain has also
an upper limit $\xi$ (see Figure~\ref{fig:fig3}) which generally depends on the sample size. For granite, the
experimental upper limit $\xi$ is of the order of one fourth of the specimen
 size~\cite{Meheust2, Auradou}.
For mortar, the geometry can be considered as self-affine only up to length scales of the order of $15\un{\%}$
of the specimen width $W$~\cite{Mourot2}. A systematic study of wood samples with various values
of $W$ leads  to $\xi=0.1 \, W$~\cite{Morel}.\\
\begin{figure}
\includegraphics[width=\W]{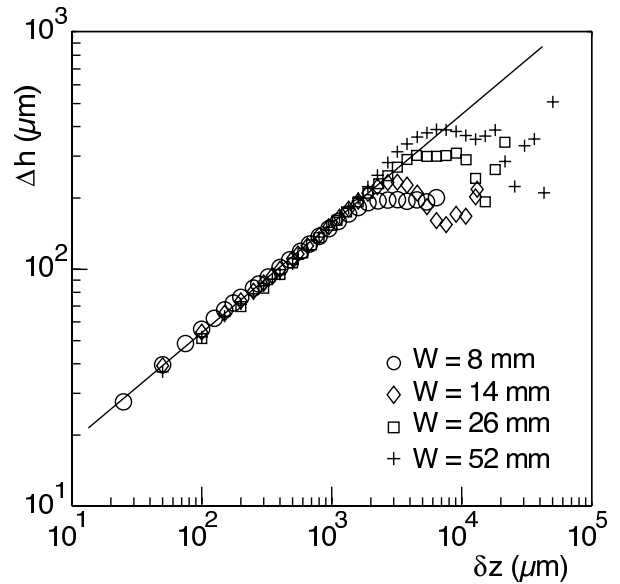}
\caption{Log-log representation of the height$-$height correlation
function $\Delta h(\delta z)$ computed along the $z$-axis for $4$ specimens of different widths $W$. The straight line is a linear regression of slope $\zeta=0.47$.} \label{fig:fig6}
\end{figure}
Figure~\ref{fig:fig6} displays  the variations of the correlation function
$\Delta h(\delta z)$ as a function of $\delta z$ for samples of different widths $W$.
All data sets display a self-affine domain followed by a plateau region corresponding
to a value of $\Delta h(\delta z>\xi)$ increasing with $W$;  the upper limit $\xi$  of the self-affine
 domain also clearly increases with  the system size.
The variation of $\xi$ with $W$ is plotted in Figure~\ref{fig:fig7} and is well fitted by a
straight line of slope  $0.15$ going through the origin.\\
\begin{figure}
\includegraphics[width=\W]{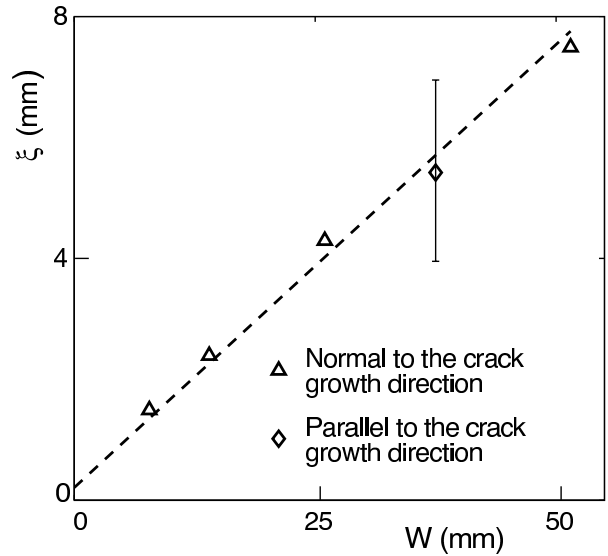}
\caption{Variations of the correlation length $\xi_z$
($\bigtriangleup$) (resp. $\xi_x$ ($\diamond$)) as a function of the specimen width (resp. length) $W$.
Slope of the dashed line:  $0.15$.} \label{fig:fig7}
\end{figure}
A value of the upper limit of the self-affine domain for profiles
parallel to the direction $x$ of crack propagation is also plotted
in Figure~\ref{fig:fig7} ($\lozenge$). This length is, in this case,
the average of values of  $\xi$ determined for four samples of different widths 
and the error bar reflects  the variability of the
values.  In  order to plot this result
 in Figure~\ref{fig:fig7}, these profiles have been assumed to
 correspond to a value of $W$ equal to  $35~mm$ (\ie the  length of the
 samples along the $x$ direction).  The corresponding data point
 is consistent with the linear fit
 corresponding in Figure~\ref{fig:fig7} to profiles parallel to the $z$-axis.
The scaling relation between $\xi$ and the system width $W$ established above
 seems apparently also applicable  in the $x$ direction.\\
Let us now compare the  surface roughness in the self-affine domain
at length scales between $d$ and $\xi$ with the anomalous scaling
characteristics recently observed~\cite{Lopez4, Morel, Mourot2}.
Anomalous growth implies that, after a transient regime, the
roughness amplitude  $\Delta h(\delta z=cst<\xi)$  scales as
$W^{\zeta_{global}-\zeta}$ with the system width $W$ (see
Eq.(\ref{eq:eq3})). All the profiles analyzed correspond to the
stationary propagation domain in order to make the comparisons
meaningful. The extent of the transient domain near the crack initiation has been found in various materials to  be of the order of few
millimeters~\cite{Ponson6,Mourot2}. 
Here, a value close to $\sim 2\un{mm}$ was measured and all profiles considered in the present work are outside this region.\\
\begin{figure}
\includegraphics[width=\W]{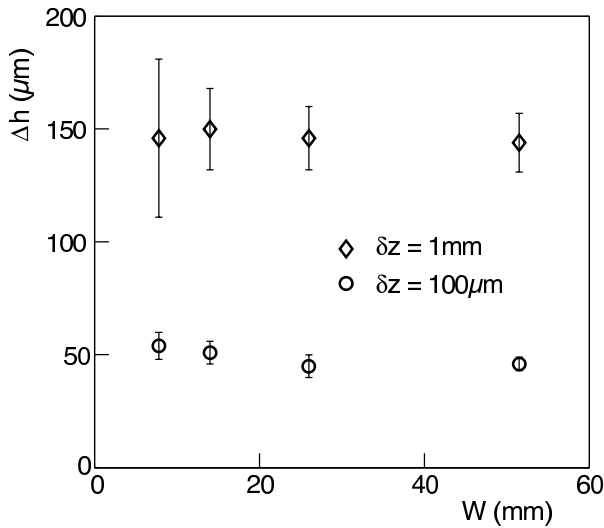}
\caption{Height$-$height correlation function $\Delta h(\delta z)$
for fixed values of $\delta z$ as a function of the specimen width
$W$. ($\diamond $): $\delta z=1\un{mm}$; ($\circ $): $\delta
z=100\unm$. Error bars: Standard deviation of $\Delta h(\delta z)$.}
\label{fig:fig8}
\end{figure}
The height$-$height correlation
function is plotted in Fig.~\ref{fig:fig8} as a function of the
specimen width $W$ for two different window sizes $\delta z$.
In contrast with the predictions of Eq.~(\ref{eq:eq3}) for anomalous scaling,
the roughness amplitude $\Delta h(\delta z<\xi)$ in the self-affine domain is independent of $W$ over a range of variations of almost $1$ to $10$ (the same conclusions may already be inferred from Fig.~\ref{fig:fig6} but the sensitivity is better here due to the linear scale).
These characteristics differ therefore completely from those expected for
anomalous scaling.

\section{Discussion and conclusions}
Let us now discuss the origin of the statistical properties observed
 for the present Fontainebleau sandstone surfaces. In a
recent study, Bonamy {\it et al.} \cite{Bonamy2} analyzed the
fracture surfaces morphology in glassy materials, \ie homogeneous
silica glass and sintered glass. They suggest that, on the one hand, 
 higher exponents $\zeta \sim 0.8$ reflect damage failure processes
occuring in a zone surrounding the crack tips and are observable at length scales smaller than the size  of the process zone; on the other hand,  low roughness
exponents $\zeta \sim 0.4$ would result from  brittle fracture  and be observable  
either in experiments where no process zone develops or at length scales larger than
 the size of the process zone. 
In this perspective,  and as is discussed in more detail  below, 
the low value  $\zeta \simeq 0.45$ reported here for 
Fontainebleau sandstone samples 
 is the signature of brittle fracture;  moreover,
since it is observed at length scales down to the grain size,  
no process zone is present in these experiments.

In spite of these important differences,  surfaces obtained from
brittle and quasi-brittle failures share  common features. In both cases, a single
exponent is enough to describe the scaling invariance of the roughness~\cite{Santucci}: this confirms the relevance of the self-affine description of fracture surfaces for both types of materials.

The lack of anomalous scaling in the present fracture surfaces compared to materials such as concrete, wood or granite displaying damage failure 
is an other indication that no damage process occurs in our experiments. 
In these latter studies, confinment  effects limit the development of the  process zone: Its extension and internal structure are then set by the system size $W$ when a  stationary propagation regime is reached. As a result, from 
Eq.~(\ref{eq:eq3})~\cite{Lopez4},  the roughness amplitude $\Delta h(\delta z=cst)$  varies with $W$ as 
$\Delta h(\delta z=cst) \propto W^{\zeta_\indice{global}-\zeta}$ as is indeed observed 
experimentally~\cite{Mourot3}. 
In contrast, in the absence of a process zone, $\Delta h(\delta z=cst)$ is only set by the
microstructural and mechanical properties of the material and does not depend on the system size (see for instance the model of Refs.~\cite{Bonamy2, Ponson10}) as observed experimentally here.

The same type of approach accounts for the domains of observation of the different self-affine geometries. For damage fractures, the upper bound $\xi$ of the self-affine domain where $\zeta \sim 0.8$ is  set by the size of the process zone~\cite{Bonamy2, Ponson5} or by the sample size  $W$ 
 if it is smaller than the process zone. 
In the present case of a brittle fracture with no process zone, the lower boundary of the self-affine domain  is  set by  the grain size.  The upper
cut-off length $\xi$ increases  linearly with the sample size $W$: in this case, no other characteristic length
than $W$ seems to determine   $\xi$ so that it  might increase indefinitely with $W$.
These results suggest that the two different self affine geometries might coexist on a same fracture surface: the characteristic exponent would have a value $\zeta \simeq 0.8$ for lengths scales up to the size of the process zone and a value $\zeta \simeq 0.4$ at larger length scales up to the plateau domain where $\Delta h(\delta z)$ becomes constant.
 
A key issue is whether the development of a process zone is purely material dependent.
Recent experiments~\cite{Backers,Backers2} suggest  that
fractured samples made of cemented grains, and therefore similar to  Fontainebleau sandstone,
  do not always display brittle failure. The
emission of acoustic waves was measured during the growth of a crack
initiated from a chevron shaped notch under fatigue in a Permian
sandstone sample. The
localization of the acoustic events during failure allowed to
demonstrate that a process zone was present and to determine its
size
 $\ell_{pz} \simeq 25\un{mm}$ and its spatio-temporal evolution~\cite{Backers2}.
Moreover, the morphology of the resulting fracture surfaces~\cite{Backers} was
 found to be self-affine for length scales less than $10 \un{mm} (< \ell_{pz})$ and characterized by an
exponent $\zeta = 0.75 \pm 0.03$, similar to the usual value for 
materials displaying damage fractures. This observation supports the assertion of Ref.~\cite{Bonamy2} and shows that the roughness exponent depends not only on the material, but also on the
experimental conditions (geometry of initiation notch, loading mode...)

To conclude, the present experimental results and their comparison to 
previous works indicate that the value of the self-affine exponent $\zeta$ is 
a clear signature of the failure mode : damage  fracture involving a 
process zone if $\zeta \simeq 0.8$ or brittle  fracture  if $\zeta \simeq 0.4$. 
Another characteristic signature of damage fracture is 
anomalous scaling marked by a variation of the roughness amplitude with the
sample size (if the latter is smaller than the process zone). 
The range of length scales over which the fracture geometry is self-affine with 
a given exponent $\zeta$ may represent an important additional information. 
For $\zeta \simeq 0.8$, this range should extend from the size  of the grains to that of the process zone (which may be limited by the sample size); for $\zeta \simeq 0.4$, it may extend from the size of the grains  (if there is no process zone) or that  of the process zone up to  a length scaling as the size of the specimen.

An important test of these hypothesis will then be to force the development 
of a process zone in materials of the type used here by modifying 
 the loading or the geometry of the experiments. For instance, one may use
an initial chevron  notch geometry as in refs.~\cite{Backers,Backers2}, where damage processes  occur (so that $\zeta = 0.8$) although the material is similar  to the Fontainbleau sandstone used here.
 If a  transition from brittle to damage fracture is induced
in this way, this would result in a variation of the self affine exponent $\zeta$
or in the coexistence of two self-affine domains with $\zeta \simeq 0.8$ below
the size of the process zone and $\zeta \simeq 0.4$ above it.

\begin{acknowledgments}
We are indebted to G. Chauvin, R. Pidoux (FAST) and J.P. Villote, V.
Godard (IDES) for their assistance in the realization of the
experimental set-up. We are grateful to E. Bouchaud, D. Bonamy, J.P.
Bouchaud, A. Hansen and K.J. M{\aa}l{\o}y for their enlightening
comments. This work is supported by the CNRS and ANDRA through the
GdR FORPRO - the EHDRA (European Hot Dry
Rock Association) - the PNRH programs and the Paris XI university
through a PPF. L. P. is partially supported by the French Ministry
of Foreign Affairs (Lavoisier Program).
\end{acknowledgments}

\end{document}